\documentclass[prd,aps,floats,twocolumn,nofootinbib]{revtex4-1}
\usepackage{slashed}
\usepackage{mathtools}
\usepackage{amsfonts}
\usepackage{amssymb}
\usepackage{amsmath}
\usepackage{comment}
\usepackage{xcolor}
\usepackage{epsfig}
\usepackage{physics}
\usepackage{epsfig}

\begin{document}

\newcommand{\m}[1]{\mathcal{#1}}
\newcommand{\nn}{\nonumber}
\newcommand{\ph}{\phantom}
\newcommand{\eps}{\epsilon}
\newcommand{\be}{\begin{equation}}
\newcommand{\ee}{\end{equation}}
\newcommand{\bea}{\begin{eqnarray}}
\newcommand{\eea}{\end{eqnarray}}
\newtheorem{conj}{Conjecture}

\newcommand{\plk}{\mathfrak{h}}
\newcommand{\bb}{\bar b}


\title{GraviGUT unification with revisited Pati-Salam model}
\date{\today}

\author{Stephon Alexander$^\dagger$, Bruno Alexandre$^*$, Michael Fine$^*$, Jo\~{a}o Magueijo$^*$, Edžus Nākums}
\email{stephon\_alexander@brown.edu}
\email{bruno.alexandre20@imperial.ac.uk}
\email{michael.fine24@imperial.ac.uk}
\email{j.magueijo@imperial.ac.uk}
\email{edzus.nakums24@imperial.ac.uk}

\affiliation{$^\dagger $Department of Physics, Brown University, Providence, RI 02912, USA}
\affiliation{$^*$Abdus Salam Centre for Theoretical Physics, Imperial College London, Prince Consort Rd., London, SW7 2BZ, United Kingdom}

\begin{abstract}
We propose a graviGUT unification scheme based on the simple orthogonal group $\mathrm{SO}(1,9     ,\mathbb{C})$ that resolves the chiral duplication of weak isospin in Pati--Salam models. In the conventional $SU(4)\times SU(2)_+\times SU(2)_-$ framework, the unobserved second chiral $SU(2)$ is typically removed by ad hoc high-energy scale breaking. Here we instead \emph{geometrize} it: one $SU(2)$ factor is identified with a chiral half of the Lorentz group, so it belongs to gravity rather than to an additional weak force. This identification becomes natural inside $\mathrm{SO}(1,9     ,\mathbb{C})$, where the algebra decomposes as $\mathfrak{so}(1,3)_{\mathbb{C}}\oplus\mathfrak{so}(6)_{\mathbb{C}}\oplus(\text{coset})$. We construct a parity-symmetric chiral action that, upon breaking \emph{dynamically} selects one chirality: the surviving Yang--Mills factor is identified with $SU(2)_+$, while the opposite chirality persists as the gravitational chiral connection.  These lead to concrete phenomenological handles, including graviton and weak-boson vertices  with the other fundamental forces in $SU(3)$ and $U(1)$ and parity-sensitive gravitational-wave signatures, that distinguish the $\mathrm{SO}(1,9,\mathbb{C})$ construction from both traditional Pati--Salam and larger, less economical unifications.
\end{abstract}

\maketitle

\section{Introduction}

The search for a common gauge geometric origin of gravity and the Standard Model has long been guided by two complementary clues: (i) gravity admits a \emph{chiral} reformulation in terms of self-dual (SD)/anti-self-dual (ASD) $SU(2)$ connections \cite{KirillBook,Freidel:2005mq,Capovilla:1991qt,Capovilla:1991kp,Plebanski:1977pw,Celada:2016jdt,Alexandre:2407.19363}, and (ii) the most economical quark--lepton unification compatible with observed quantum numbers is the \emph{Pati--Salam} group $SU(4)\otimes SU(2)_L\otimes SU(2)_R$ \cite{PatiSalam1974}. In this work we argue that a unification based on the simple orthogonal group $SO(10,\mathbb{C})$ (in the form $SO(1,9, \mathbb{C})$ gauged over 4D Lorentzian spacetime) marries these clues in an especially natural way. The algebra splits as
\begin{equation}\label{split}
    \mathfrak{so}(1,9)_{\mathbb{C}}\;\simeq\;\mathfrak{so}(1,3)_{\mathbb{C}}\oplus\mathfrak{so}(6)_{\mathbb{C}}\oplus(\text{coset}),
\end{equation}
with $\mathfrak{so}(6)\cong\mathfrak{su}(4)$ providing the Pati--Salam $SU(4)$ part that is aligned with its observational benefits, while $\mathfrak{so}(1,3)_{\mathbb{C}}\cong \mathfrak{sl}(2,\mathbb{C})_+\oplus\mathfrak{sl}(2,\mathbb{C})_-$ accounts for the chiral Lorentz sector and the weak force upon applying appropriate reality conditions. This already hints at a structural identification between one Pati--Salam $SU(2)$ factor and a real chiral half of the Lorentz group. With this in mind, let us now analyse
the choice of the Pati-Salam model.

Conceptually, the Pati–Salam framework \(\big(SU(4)\times SU(2)_+\times SU(2)_-\big)\) brings a number of virtues we retain. An example is that $SU(4)$ unifies quarks and leptons by treating lepton number as a fourth colour, which naturally explains charge quantization and gives a gauge origin to \(B\!-\!L\) \cite{PatiSalam1974}. It also embeds neatly inside \(SO(10)\) \cite{BarrRaby1997}, providing a coherent path to full unification with minimal matter multiplets. 
However, the model's core phenomenological drawback lies in its extra \(SU(2)_-\). This might give us  the benefit of a chiral symmetric theory, but unfortunately it also predicts right-handed charged currents and new gauge bosons (\(W_R\), often accompanied by a \(Z'\)) that have not been observed—current LHC analyses exclude \(W_R\) up to about \(6.4\ \text{TeV}\) under standard assumptions \cite{ATLAS2023WR}. Lack of observation of these phenomena pushes the \(SU(2)_-\) breaking scale high and constrains viable realizations, as emphasized throughout the left–right symmetric literature \cite{MohapatraSenjanovic1975,Raby2006}.

A key technical reason for choosing $SO(1,9, \mathbb{C})$ over larger simple groups (e.g.\ $SO(3,11)$ or $E_8$) is that $SO(1,9, \mathbb{C})$ is the smallest simple group that contains $\mathfrak{so}(1,3)_{\mathbb{C}}\oplus\mathfrak{so}(6)_{\mathbb{C}}$ as a subalgebra and whose coset directions carry the correct bifundamental indices to play the role of a \emph{vierbein} \cite{followup}. 
This converts a unified curvature squared invariant into the Einstein--Cartan term plus a cosmological volume form without explicitly breaking the residual internal gauge symmetry. The same ansatz generates, at the same order, conventional Yang--Mills kinetic terms for the internal sector and torsion dynamics, all descending from the an $SO(1,9, \mathbb{C})$ gauge invariant action. 
So the Lorentz $SU(2)$ is not an additional high--energy gauge sector, but rather is part of the gravitational sector \cite{Alexander:2012ge,Alexander:0706.4481,Nesti:0706.3307,Nesti:0706.3304,Alexander:2011qfa}; this explains why we do not observe a second weak interaction in the infrared.

Beyond kinematics, we show that parity can be made a fundamental symmetry of the high--energy action and then broken \emph{dynamically} so that one chiral sector reduces to a purely Yang-Mills action for the weak force and the other to a self-dual action for gravity. This is achieved by using a pseudo-scalar Higgs field \cite{higgs} transforming oddly under $Z_2$, that breaks parity upon fixing a vacuum expectation value, and a second $SO(1,9,\mathbb{C})$ Higgs field which spontaneously breaks the symmetry down to $SO(1,6,\mathbb{C})\times SO(1,3,\mathbb{C})$. 

Finally, regarding phenomenology, this framework has resultant coupling terms which imply graviton and weak--boson vertices with the strong and electromagnetic forces. These would have parity--sensitive signatures in the gravitational wave background--clean phenomenological targets that distinguish the $SO(1,9, \mathbb{C})$ construction from both minimal gravi-weak and traditional grand unification.

The plan of this paper is as follows: in Section \ref{GWHMM} we start by laying out the theory, introducing the $SO(1,9,\mathbb{C})$ connection and deriving the corresponding fields strength components. We further introduce the action of the theory, together with the algebra simplifications and Higgs mechanisms that lead to the final graviGUT Lagrangian. In Section \ref{su4} we introduce a map from $SO(6)$ to $SU(4)$ in order to further decompose into $SU(3)\times U(1)$, the groups that represent the strong and electromagnetic (EM) forces. Section \ref{spurious} is dedicated to a brief analysis of the degrees of freedom encoded in the coset part of the $SO(1,9,\mathbb{C})$ connection. Finally, we close with some concluding remarks and an outlook of future work in Section \ref{conc}.

\section{Modified Pati-Salam $\boldsymbol{SO(1,9, \mathbb{C})}$ model}\label{GWHMM}
In this Section we set up the basic structure of the modification of the Pati-Salam model we wish to propose.

\subsection{Structure of the Gauge Fields}\label{GaugeFields}
The model is based on the decomposition:
\begin{eqnarray}
    \mathfrak{so}(1,9)_{\mathbb{C}} \;=\; 
\underbrace{\mathfrak{so}(1,3)_{\mathbb{C}}}_{(\mathrm{Adj},\,\mathbf{1})}
\;\oplus\;
\underbrace{\mathfrak{so}(6)_{\mathbb{C}}}_{(\mathbf{1},\,\mathrm{Adj})}
\;\oplus\;
\underbrace{\big(\mathbb{C}^{1,3}\otimes \mathbb{C}^{6}\big)}_{(\mathbf{4},\,\mathbf{6})}. \label{eqgroupsplit}
\end{eqnarray}
The generators  of $\mathfrak{so}(1,9)_{\mathbb{C}}$, $M_{AB}$, obey the commutation relation $
[M_{AB}, M_{CD}] = \eta_{AD} M_{BC} - \eta_{AC} M_{BD} - \eta_{BD} M_{AC} + \eta_{BC} M_{AD}
$
and have indices running as $A,B,C...=a,b,c,...,I,J,K,...=0,...,9$, where $a,b,c,..=0,1,2,3$ are in \( \mathfrak{so}(1,3)_{\mathbb{C}} \) and $I,J,K,...=4,...,9$ in \( \mathfrak{so}(6)_{\mathbb{C}} \). We therefore have $45$ generators of $SO(1,9,\mathbb{C})$, which break into the 6 generators of $SO(1,3,\mathbb{C})$ plus the $15$ generators of $SO(6,\mathbb{C})$ and $24$ of the coset $\big(\mathbb{C}^{1,3}\otimes \mathbb{C}^{6}\big)$.
The gauge field is then an \( \mathfrak{so}(1,9)_{\mathbb{C}} \) connection with components \( \mathcal{A}^{AB} \). Its field strength is given by:
\[
\mathcal{F} = d\mathcal{A} + \mathcal{A} \wedge \mathcal{A}=\frac{1}{2}\mathcal{F}_{AB}M^{AB},
\]
with components
$\mathcal{F}_{AB} = d\mathcal{A}_{AB} + \mathcal{A}_{AC} \wedge \mathcal{A}^C_{~B}$. In line with the split (\ref{eqgroupsplit}) we decompose:
\begin{equation}
    \mathcal{A}=\omega+A+\phi
\end{equation}
where $\omega=\frac{1}{2}\omega_{ab}M^{ab}$ is the $SO(1,3,\mathbb{C})$ connection, $A=\frac{1}{2}A_{IJ}M^{IJ}$ is the $SO(6,\mathbb{C})$ connection and $\phi=\phi_{aI}M^{aI}$ the coset field. The total curvature components  $\mathcal{F}_{AB}$ can then be written as: 
\begin{eqnarray}
&&\mathcal{F}^a_{~b}=R^a_{~b}+\phi^a_{~I}\wedge \phi^I_{~b}, \label{eqcurvatures1}\\
    && \mathcal{F}^I_{~J}=F^I_{~J}+\phi^I_{~a}\wedge \phi^a_{~J}, \label{eqcurvatures2}\\
    && \mathcal{F}^a_{~I}=d\phi^a_{~I}+\omega^a_{~c}\wedge\phi^c_{~I}+\phi^a_{~J}\wedge A^J_{~I}.
    \label{eqcurvatures3}
\end{eqnarray}
where $R^{ab}=d\omega^{ab}+\omega^a{}_c\wedge\omega^{cb}$ and $F^I{}_J=dA^{I}{}_{J}+A^I{}_K\wedge A^{K}{}_{J}$.

We now consider the symmetry breaking McDowell--Mansouri--like ansatz \cite{MacDowellMansouri}:
\begin{eqnarray}
    && \phi^{aI} = \frac{1}{\ell} e^a n^I, 
    \label{mcdowell-M} \\
    && \phi^{Ia} = -\frac{1}{\ell} e^a n^I,
\end{eqnarray}
where $e^a$ is the tetrad, $\ell$ is a constant, and $n^I$ is a normalized internal vector $(n^2=1)$. This reduces the number of independent coset components from $6\times 4\times4 = 96$ to $6+(4\times4)=22$, and we will revisit this condition later in this paper.
Equations (\ref{eqcurvatures1}) and (\ref{eqcurvatures2}) then become
\begin{eqnarray}
    && \mathcal{F}^a_{~b}=R^a_{~b}-\ell^{-2}e^a\wedge e_b, \label{eqfab} \\
    && \mathcal{F}^I_{~J}=F^I_{~J} \label{eqfij}
\end{eqnarray}
and for the field strength in the off diagonal block one gets
\begin{equation}
\mathcal{F}^{aI} = \frac{1}{\ell} \big(de^{a} + \omega^{a}{}_{b} \wedge e^{b} \big) n^{I}
+ \frac{1}{\ell} e^{a}\wedge\big( dn^{I} + A_{J}{}^{I} n^{J} \big). 
\end{equation}
The first term depends on the torsion of the spacetime geometry $T^a$, while the second term depends on the internal dynamics of the vector $n^{I}$.
Torsion is defined in the usual way and we can introduce a new covariant object $P^I$, analogous to torsion but in the internal $SO(6,\mathbb{C})$ space. Hence:
\begin{equation}
T^{a} \equiv D e^{a} = de^{a} + \omega^{a}{}_{b} \wedge e^{b}, 
\quad 
P^{I} \equiv D n^{I} = dn^{I} + A_{J}{}^{I} n^{J}.
\end{equation}
In terms of these quantities the off diagonal curvature becomes
\begin{equation}
\mathcal{F}^{aI} = \frac{1}{\ell}( T^{a} n^{I} +  e^{a}\wedge P^{I}) .\label{eqfai}
\end{equation}

\subsection{The action }

With the field strength specified by our ansatz, we now construct an action subject to the Graviweak criteria \cite{followup}: (i) diffeomorphism covariance; (ii) parity covariance, realised as spatial inversion accompanied by an exchange of chiral sectors; and (iii) locality, restricting terms to be at most linear or quadratic in the curvature two–forms. The action is built from the curvature \(\mathcal{F}^{AB}\) of the \(SO(1,9,\mathbb{C})\) connection, an \(SO(1,9,\mathbb{C})\) Higgs field \(U^{A_1\cdots A_5}\), and the real pseudoscalar \(\phi\) ensuring parity invariance:
\begin{widetext}
\begin{eqnarray}
   && S=\frac{\alpha}{2}\int\mathcal{F}^{AB}\wedge*\mathcal{F}_{AB}+\frac{\gamma\alpha}{4}\int \phi\epsilon_{A_1...A_{10}}\mathcal{F}^{A_1 A_2} \wedge*{\mathcal{F}}^{A_3 A_4} U^{A_5...A_{10}}+\nonumber\\
&&+\gamma\alpha\int\phi\mathcal{F}^{AB}\wedge\mathcal{F}_{AB}+\int [V(\phi)+V(U)]*1, \label{eqfulls}
\end{eqnarray} 
\end{widetext}
where $*$ is the Hodge dual in spacetime, $\gamma$ is an Immirzi-like parameter and $\alpha$ is a coupling constant. One can also interpret $\gamma \phi$ as a dynamical Immirzi-like parameter. The first term is a kinetic term for the curvature, the second is a Stelle-West-like term and the third is a Pontryagin-like term. \(V(\phi) \) is the Higgs potential the pseudoscalar field that transforms under parity as $\phi'=-\phi$,  
\[ V(\phi) = \frac{\lambda}{4} (\phi^2 - 1)^2, \] 
leading to spontaneous symmetry breaking with vacuum expectation values (vev) \( \langle \phi \rangle = \pm 1 \), and where \( \lambda \) is a coupling constant. $V(U)$ is the Higgs potential for the field $U^{A_1...A_6}$:
\begin{eqnarray}
    V(U)=\frac{\beta}{4}(U_{A_5...A_{10}}U^{A_5...A_{10}}-1)^2, 
    \label{UVev}
\end{eqnarray}
where $\beta$ is a coupling constant. In the unitary gauge we take a vev
\(U_0^{I_1\ldots I_6}=\frac{1}{6!}\epsilon^{I_1\ldots I_6}\)
(so \(\,U_0\cdot U_0=1\)),
which spontaneously breaks \(SO(1,9,\mathbb C)\) into the internal $SO(6,\mathbb C)$ block and the complex Lorentz $SO(1,3,\mathbb C)$ block; here
\(\epsilon^{I_1\ldots I_6}\) is the Levi–Civita tensor of the internal \(SO(6,\mathbb C)\).

We will now work though the action term by term.
The first can be explicitly written in terms of the subgroups and coset indices:
\begin{eqnarray}
&&\mathcal{F}^{AB} \wedge *\mathcal{F}_{AB}=\mathcal{F}^{ab} \wedge *\mathcal{F}_{ab}+2\mathcal{F}^{aI} \wedge *\mathcal{F}_{aI}+\nonumber\\
&&+\mathcal{F}^{IJ} \wedge *\mathcal{F}_{IJ}.\label{eqfsf}
\end{eqnarray}
Upon inserting equation (\ref{eqfai}) the term $\mathcal{F}^{aI} \wedge *\mathcal{F}_{aI}$ expands as
\begin{eqnarray}
    && \mathcal F^{aI} \wedge *\,\mathcal F_{aI}
=\frac{1}{\ell^{2}}\, T^{a} \wedge * T_{a}
\;-\frac{1}{\ell^{2}}\, T^{a} n^{I} \wedge * (P_{I} e_{a})
+\nonumber\\
&& -\frac{1}{\ell^{2}}\, P^{I} e^{a} \wedge * T_{a} n_{I}
\;+\; \frac{1}{\ell^2}P^{I} e^{a} \wedge * (P_{I} e_{a}). \nonumber 
\end{eqnarray}
Moreover, \(T^{a} n^{I} \wedge * (P_{I} e_{a})=P^{I} e^{a} \wedge * T_{a} n_{I}=0\) since \(n_{I} P^{I}
= n_{I}\, d n^{I} + n_{I} A^{I}{}_{J} n^{J}
= \tfrac12\, d(n_{I} n^{I})= \tfrac12\, d(n^{2})=0\).
The last term of this expansion can be simplified to
\[
e_{a} \wedge * (P_{I} e^{a})= 3\,* P_{I},
\]
which leads to
\[
\Rightarrow\qquad
P^{I} e^{a} \wedge * (P_{I} e_{a})
\;=\; 3\, P^{I} \wedge * P_{I}.
\]
Hence we have:
\begin{eqnarray}
    \mathcal{F}^{aI} \wedge *\mathcal{F}_{aI}=\frac{1}{\ell^2}T^a\wedge*T_a+ \frac{3}{\ell^2}P^I\wedge*P_I.
\end{eqnarray}
When the field $U$ condensates at its vev, the second term in (\ref{eqfulls}) becomes 
\begin{eqnarray} \epsilon_{A_1...A_{10}}\mathcal{F}^{A_1 A_2} \wedge*{\mathcal{F}}^{A_3 A_4} U^{A_5...A_{10}}=\epsilon_{abcd}\mathcal{F}^{ab}\wedge*\mathcal{F}^{cd}.\nonumber
\label{interaction1}
\end{eqnarray}
Furthermore, the expansion around the vev, $U=U_0+\delta U$, leads to very interesting interactions between gravity, with the $SO(6)$ block (which encodes the strong force and EM), and likewise between the weak force, and the $SO(6)$ block but their detailed analysis is beyond the scope of this work. These terms are of the form
\begin{eqnarray}
    \label{interaction2}
    && \epsilon_{abcdIJKLMN}\mathcal{F}^{ab}\wedge*\mathcal{F}^{IJ}\delta U^{cdKLMN} \\
    \label{interaction3}
    && \epsilon_{IJKLMNabcd}\mathcal{F}^{IJ}\wedge*\mathcal{F}^{KL}\delta U^{MNabcd} \\
    && \epsilon_{IJaKLMNbcd}\mathcal{F}^{IJ}\wedge*\mathcal{F}^{aK}\delta U^{LMNbcd} \\
    && \epsilon_{aIbJcdKLMN}\mathcal{F}^{aI}\wedge*\mathcal{F}^{bJ}\delta U^{cdKLMN} \\
    && \epsilon_{abcIdJKLMN}\mathcal{F}^{ab}\wedge*\mathcal{F}^{cI}\delta U^{dJKLMN}.
    \label{endinteractions}
\end{eqnarray}

Upon spontaneous breaking in the pseudoscalar sector we set \(\langle \phi \rangle = 1\). With the parameter choice \(\gamma = i\), the interaction term \eqref{interaction2} recombines with the leading YM term in \eqref{eqfsf}, yielding their chiral combination:
\begin{eqnarray}
    && \frac{1}{2}\mathcal{F}^{ab} \wedge *\mathcal{F}_{ab}+\frac{i}{4}\epsilon_{abcd}\mathcal{F}^{ab}\wedge*\mathcal{F}^{cd}=\mathcal{F}^{(+)ab} \wedge *\mathcal{F}^{(+)}_{ab}=\nonumber \\
    && =R^{(+)ab} \wedge *R^{(+)}_{ab}-\frac{2}{\ell^{2}}\, \Sigma^{(+)}_{ab} \wedge *R^{(+)ab}+\nonumber\\
    &&+\frac{1}{\ell^{4}} \Sigma^{(+)}_{ab}\,  \wedge*\Sigma^{(+)ab}\label{eqf+},
\end{eqnarray}
where we introduced $\Sigma^{ab}=e^{a} \wedge e^{b}$, and we also used equation (\ref{eqfab}) and the projectors to write the ASD($+$)/SD($-$) components:
\begin{eqnarray}
    P^{(\pm)ab}{}_{cd}=\frac{1}{2}\left(\delta^{ab}_{cd}\pm\frac{i}{2}\epsilon^{ab}{}_{cd}\right).
    \label{Projectors}
\end{eqnarray}
To further decompose equation (\ref{eqf+}) in terms of the SD and ASD components we start by working out the second term explicitly:
\begin{eqnarray}
    && \Sigma^{(+)}_{ab} \wedge *R^{(+)ab}=\frac{1}{2}\epsilon_{abcd}R^{(+)ab}\wedge \Sigma^{(+)cd}\nonumber\\&&=-iR^{(+)ab}\wedge \Sigma^{(+)}_{ab}. 
\end{eqnarray}
Finally, the last term yields
\begin{eqnarray}
   && \Sigma^{(+)}_{ab}\,  \wedge*\Sigma^{(+)ab}=-i\Sigma^{(+)ab}\wedge\Sigma^{(+)}{}_{ab}=i\Sigma^{(-)ab}\wedge\Sigma^{(-)}{}_{ab},\nonumber
\end{eqnarray}
where we note that the cosmological constant term can be reformulated in this way. 
\newline
Putting these together, equation (\ref{eqf+}) takes the form
\begin{widetext}
\begin{eqnarray}
    \frac{1}{2}\mathcal{F}^{AB} \wedge *\mathcal{F}_{AB}+\frac{i}{4}\epsilon_{abcd}\mathcal{F}^{ab}\wedge*\mathcal{F}^{cd}&=&R^{(+)ab} \wedge *R^{(+)}_{ab}+\frac{i}{\ell^4}\Sigma^{(-)ab}\wedge\Sigma^{(-)}{}_{ab}+\frac{2i}{\ell^2}R^{(+)ab}\wedge \Sigma^{(+)}_{ab}\nonumber\\
    &+&\frac{1}{\ell^2}T^a\wedge*T_a+ \frac{3}{\ell^2}P^I\wedge*P_I+\frac{1}{2}F^{IJ}\wedge*F_{IJ}
    \label{fpsfp}
\end{eqnarray}
\end{widetext}

Finally, let us analyse the $\mathcal{F}\wedge\mathcal{F}$ term in (\ref{eqfulls}), which can be written explicitly as
\begin{eqnarray}
&&\mathcal{F}^{AB} \wedge \mathcal{F}_{AB}=\mathcal{F}^{ab} \wedge \mathcal{F}_{ab}+2\mathcal{F}^{aI} \wedge \mathcal{F}_{aI}+\nonumber\\
&&+\mathcal{F}^{IJ} \wedge \mathcal{F}_{IJ},
\end{eqnarray}
where by mirroring the procedure done for the kinetic term one obtains
\begin{eqnarray}
    2\mathcal{F}^{aI} \wedge \mathcal{F}_{aI}+\mathcal{F}^{IJ} \wedge \mathcal{F}_{IJ}=\frac{2}{\ell^2}T^a\wedge T_a+F^{IJ}\wedge F_{IJ}, \nonumber
\end{eqnarray}
and consequently,
\begin{eqnarray}
    && \mathcal{F}^{ab} \wedge \mathcal{F}_{ab}=R^{ab} \wedge R_{ab}-\frac{2}{\ell^{2}}\, e_{a} \wedge e_{b} \wedge R^{ab}= \nonumber \\
    && =R^{ab} \wedge R_{ab}-\frac{2}{\ell^{2}}R^{(+)ab}\wedge \Sigma^{(+)}_{ab}-\frac{2}{\ell^{2}}R^{(-)ab}\wedge \Sigma^{(-)}_{ab}.\nonumber
\end{eqnarray}
Having worked out all the terms in the full action, the Lagrangian density \eqref{eqfulls} can be recast as
\begin{widetext}
\begin{eqnarray}
\nonumber\mathcal{L}&=&\frac{-2i \alpha }{\ell^{2}} \Sigma_{ab}^{(-)} \wedge R^{(-)ab}+\frac{i\alpha }{\ell^{4}}\Sigma^{(-)ab} \wedge \Sigma^{(-)}_{ab}+\\&&+{\alpha}\left(R_{ab}^{(+)}\wedge *R^{(+)ab}+\frac{1}{2}F^{IJ}\wedge*F_{IJ}+\frac{1}{\ell^2}T^a\wedge*T_a + \frac{3}{\ell^2}P^I\wedge*P_I \right )+\nonumber\\
&&+i\alpha\left( R^{ab} \wedge R_{ab}+F^{IJ}\wedge F_{IJ}+\frac{2}{\ell^2}T^a\wedge T_a\right).
\label{fulldecomposition}
 \end{eqnarray}
 \end{widetext}
The constants $\alpha$ and $\ell$ are then chosen in order to return the correct prefactors on the gravitational and cosmological constant terms, enforcing the following identification:
\begin{eqnarray}
\alpha &=& -\frac{1}{32\pi G}\,\frac{3}{2\Lambda}\,,\qquad
\frac{1}{\ell^{2}} \;=\; \frac{2\Lambda}{3}\,.
\label{constants}
\end{eqnarray} 
After integrating out topological terms, disregarding torsion, and restoring the action in terms of the tetrad in the familiar EC formulation, we have the final GraviGUT Lagrangian, which is the central result of this paper:
\begin{widetext}
\begin{eqnarray}
\nonumber\mathcal{L}&=&\frac{1}{32\pi G} \epsilon_{abcd}\left(e^a\wedge e^b \wedge R^{(-)cd}-\frac{\Lambda}{6}e^a\wedge e^b \wedge e^c\wedge e^d  \right)-\frac{3}{32\pi G}   P^I\wedge*P_I+\\&&-\frac{3}{64\pi G\Lambda}\left( R_{ab}^{(+)}\wedge *R^{(+)ab}+\frac{1}{2}F^{IJ}\wedge*F_{IJ}\right).
\label{fulldecomposition2}
\end{eqnarray}
\end{widetext}

To finally identify the chiral connections with gravity and the weak force, one imposes the same reality conditions mentioned as \cite{followup,Jacobson:1988fb}. In addition one also requires the internal $SO(6)$ connection $A^{IJ}$ to be real.

The term-by-term interpretation of equation \eqref{fulldecomposition2} goes as follows:
the first two terms are the familiar Einstein-Cartan plus cosmological constant Lagrangian which describe standard GR. The term \(\tfrac{3}{32\pi G}\,P^{I}\wedge *P_{I}\) is a quadratic “kinetic” contribution from the internal one-form (the projected coset field). 
Note that in this case, the cosmological constant term naturally arises, yet unlike Stelle-West, we cannot link it to any dS or AdS groups. Finally, the terms \(\big({\alpha}R^{(+)}_{ab}\wedge *R^{(+)ab}+\frac{\alpha}{2} F^{IJ}\wedge *F_{IJ} \big)\) contain curvature squared terms for the $SU(2)_R$ and for the internal \(SO(6)\) field strengths, and correspond to the kinetic terms of these fields.

We can also consider two other alternatives to derive an equivalent GraviGUT Lagrangian. One is to replace the $\gamma\phi\mathcal{F}^{AB}\wedge \mathcal{F}_{AB}$ term in (\ref{eqfulls}) with 
\begin{eqnarray}
    -\frac{1}{2} \epsilon_{A_1...A_{10}}\mathcal{F}^{A_1 A_2} \wedge{\mathcal{F}}^{A_3 A_4} U^{A_5...A_{10}}=-\frac{1}{2} \epsilon_{abcd}\mathcal{F}^{ab}\wedge\mathcal{F}^{cd},\nonumber
\end{eqnarray}
which reads
\begin{widetext}
\begin{eqnarray}
    -\frac{1}{2} \epsilon_{abcd}\mathcal{F}^{ab}\wedge\mathcal{F}^{cd}=-\frac{1}{2}\epsilon_{abcd}R^{ab}\wedge R^{cd}-\frac{2i}{\ell^2}\left(R^{(+)ab}\wedge \Sigma^{(+)}_{ab}-R^{(-)ab}\wedge \Sigma^{(-)}_{ab}
    \right)-\frac{2i}{\ell^4}\Sigma^{(-)ab}\wedge\Sigma^{(-)}_{ab}.
\end{eqnarray}
\end{widetext}
The first term is a topological term which can be integrated out, and the remaining terms can be combined with equation (\ref{fpsfp}) to yield the same GraviGUT Lagrangian as in (\ref{fulldecomposition2}), but with constants $\alpha$ and $\ell^2$ given by
\begin{eqnarray}
\alpha &=& -\frac{1}{32\pi G}\,\frac{3}{2\Lambda}\,,\qquad
\frac{1}{\ell^{2}} \;=-\; \frac{2\Lambda}{3}\,.
\label{constants2}
\end{eqnarray} 

Finally, one can consider a third alternative action which includes all possible invariant combinations built from the field strength and the Higgs fields. For instance, we can write:
\begin{widetext}
\begin{eqnarray}
   && S=\int\frac{\alpha}{2}\mathcal{F}^{AB}\wedge*\mathcal{F}_{AB}+\frac{\gamma\alpha}{4} \phi\epsilon_{A_1...A_{10}}\mathcal{F}^{A_1 A_2} \wedge*{\mathcal{F}}^{A_3 A_4} U^{A_5...A_{10}}+\nonumber\\
&&+2\gamma\int \frac{\alpha}{2}\phi\mathcal{F}^{AB}\wedge\mathcal{F}_{AB}+\frac{\gamma\alpha}{4} \epsilon_{A_1...A_{10}}\mathcal{F}^{A_1 A_2} \wedge{\mathcal{F}}^{A_3 A_4} U^{A_5...A_{10}}+\int [V(\phi)+V(U)]*1, \label{eqfulls}
\end{eqnarray} 
\end{widetext}
where every quantity was previously introduced. 

The linear combination of the bottom two terms yields:
\begin{widetext}
\begin{eqnarray}
    \frac{1}{2}\mathcal{F}^{AB} \wedge \mathcal{F}_{AB}+\frac{i}{4}\epsilon_{abcd}\mathcal{F}^{ab}\wedge\mathcal{F}^{cd}&=&R^{(+)ab} \wedge R^{(+)}_{ab}-\frac{1}{\ell^4}\Sigma^{(-)ab}\wedge\Sigma^{(-)}{}_{ab}-\frac{2}{\ell^2}R^{(+)ab}\wedge \Sigma^{(+)}_{ab}\nonumber\\
    &+&\frac{1}{\ell^2}T^a\wedge T_a+\frac{1}{2}F^{IJ}\wedge F_{IJ},
    \label{fpsfp2}
\end{eqnarray}
\end{widetext}
and integrating by parts the torsion term, \(T^a\wedge T_a=R_{ab}\wedge\Sigma^{ab}\) (disregarding the boundary term),together with all the previous simplifications for the remaining terms of the action, one can recover once again the final GraviGUT Lagrangian (\ref{fulldecomposition2}).

These identifications have an immediate physical interpretation. The YM coupling in
\begin{equation}
    \mathcal L_{\rm YM}\supset \alpha\Big(\,R^{(+)}_{ab}\wedge*R^{(+)ab} + \frac{1}{2}F^{IJ}\wedge *F_{IJ}\Big),
\end{equation}
can be identified as $\alpha = -\frac{1}{8{g_W}^2}$. If $\ell$ is then identified with the Planck length $\ell \sim \ell_\text{Pl} = \sqrt{8\pi G}$, then $g_W \sim 0.707$. This is remarkably close to the observed electroweak value $g_2 \sim 0.65$.

This is remarkably close to the observed electroweak value $g_2\sim 0.65$.
(Indeed, evolving $g$ from the high scale $\mu_0=\bar M_{\rm Pl}$ to  boson scale $M_{z}$ with one-loop RGEs—
taking $b_2^{\rm MSSM}=+1$ above $M_{\rm SUSY}\sim 1~\mathrm{TeV}$, with

$\displaystyle \frac{1}{g^2(\mu)}=\frac{1}{g^2(\mu_0)}-\frac{b_2}{8\pi^2}\ln\!\frac{\mu}{\mu_0}$ 
gives $\displaystyle g_2(M_Z)\simeq 0.65$. \cite{Schwartz2014}\cite{ChengLi1984})

At the same time, these choices fix what we will call the bare cosmological constant via $\Lambda_\text{bare} = -\frac{3}{\ell^2}$ which is to be distinguished from the effective $\Lambda$. Thus, a single geometric length controls both the normalisation of gauge-kinetic terms and the cosmological constant, tying the strength of the weak interaction directly to the microscopic soldering scale of spacetime. The tetrad's intrinsic length acts as a genuine physical scale linking gravity and the weak / Pati-Salam sectors, with renormalisation group running supplying the bridge to low-energy data.

\section{SU(4) Basis}
\label{su4}
We now introduce a map between $SO(6)$ and $SU(4)$ since the latter is part of the Pati-Salam group and upon symmetry breaking contains $SU(3)\times U(1)$, which should represent the strong and EM interactions, respectively.
To do so, we use the local Lie algebra isomorphism $\,\mathfrak{so}(6)\cong\mathfrak{su}(4)\,$. A concrete bridge between the two is provided by a Clifford algebra realisation of $\mathfrak{so}(6)$. Let $\Gamma^{I}$ be $6$ gamma matrices obeying the Clifford relation
\begin{equation}
\{\Gamma^{I},\Gamma^{J}\}=2\,\delta^{IJ},
\end{equation}
and define the generators
\begin{equation}
\sigma^{IJ} \;=\; \tfrac14[\Gamma^{I},\Gamma^{J}].
\end{equation}
The matrices $\sigma^{IJ}$ satisfy the $\mathfrak{so}(6)$ commutation relations and therefore construct a faithful representation of the algebra. Acting on the chiral spinor representation, these same generators correspond to $\mathfrak{su}(4)$, making the isomorphism explicit at the level of matrices.
With this in hand, the internal gauge connection can be written in either basis without ambiguity. If $T_{i}$, $i=1,\dots,15$, denotes a basis of $\mathfrak{su}(4)$ then the same $1$–form connection admits the two equivalent expansions
\begin{equation}
A \;=\; \tfrac12\,A^{IJ}\,\sigma_{IJ} \;=\; A^{i}\,T_{i}.
\end{equation}
Consequently, the same dictionary applies to the curvature:
\begin{equation}
F = dA + A\wedge A = F^{i}\,T_{i},
\end{equation}
with
\begin{eqnarray}
    F^{i} = dA^{i} + \tfrac12\,f^{i}{}_{jk}\,A^{j}\wedge A^{k}.
\end{eqnarray}
We then decompose the adjoint of $SU(4)$ along the subgroup $SU(3)\times U(1)$. At the level of representations one has $\,\mathbf{15}\to\mathbf{8}\oplus\mathbf{1}\oplus\mathbf{6}\,$, where $\mathbf{8}$ is the adjoint of $SU(3)$, $\mathbf{1}$ the trivial representation, and $\mathbf{6}$ the coset. Choosing an index convention in which $\hat{i}=1,\dots,8$ labels the $SU(3)$ generators, $15$ labels the $U(1)$ generator, and $\hat{a}=1,\dots,6$ the coset directions. The connection components are grouped as
\begin{equation}
A^{i} = \bigl(\,W^{\hat{i}}, \psi^{\hat{a}}, B\,\bigr).
\end{equation}
The corresponding curvatures follow from $F^{i}=dA^{i}+\tfrac12 f^{i}{}_{jk}A^{j}\wedge A^{k}$. The $SU(3)$ part is the usual non–Abelian field strength, now with a quadratic coset term,
\begin{equation}
F^{\hat{i}}
\;=\;
dW^{\hat{i}} + \tfrac12\,f^{\hat{i}}{}_{\hat{j}\hat{k}}\,W^{\hat{j}}\wedge W^{\hat{k}}
\;+\;\tfrac12\,f^{\hat{i}}{}_{\hat{a}\hat{b}}\,\psi^{\hat{a}}\wedge\psi^{\hat{b}}.
\end{equation}
The $U(1)$ curvature is Abelian but can acquire a coset bilinear through the non–trivial $U(1)$ charges of the coset,
\begin{equation}
F^{15}
\;=\;
dB
\;+\;\tfrac12\,f^{15}{}_{\hat{a}\hat{b}}\,\psi^{\hat{a}}\wedge\psi^{\hat{b}}.
\end{equation}
Finally, the coset curvature couples to both subalgebras,
\begin{equation}
F^{\hat{a}}
\;=\;
d\psi^{\hat{a}}
\;+\;\tfrac12\,f^{\hat{a}}{}_{\hat{b}\hat{c}}\,\psi^{\hat{b}}\wedge\psi^{\hat{c}}
\;+\;f^{\hat{a}}{}_{\hat{b}\hat{i}}\,\psi^{\hat{b}}\wedge A^{\hat{i}}
\;+\;f^{\hat{a}}{}_{15\hat{b}}\,A^{15}\wedge\psi^{\hat{b}}.
\end{equation}
The structure constants appearing here reflect the chosen embedding. In particular, 
\begin{equation*}
    f^{15}{}_{ij} \;=f^{a}{}_{bc} \;=\; f^{a}{}_{ij}
        \;=\; f^{15}{}_{a i}\;=\;0.
        \qquad
\end{equation*}
The purely $SU(3)$ constants $f^{\hat{i}}{}_{\hat{j}\hat{k}}$ take their standard values, while $f^{\hat{i}}{}_{\hat{a}\hat{b}}$ encode how the coset transforms as an $SU(3)$ multiplet. With this in mind and all forces now being identified within the full $SO(1,9,\mathbb{C)}$ group, equations
(\ref{interaction1}),
(\ref{interaction2}), and  (\ref{interaction3}) clearly describe interaction terms between all the fundamental forces with each other, bar gravity and the weak force as the chiral duals will terminate each other. These terms will be discussed further in outlook. 

\section{Analysis of Spurious Degrees of Freedom}
\label{spurious}
The coset ansatz introduced in (\ref{mcdowell-M}) reduces the degrees of freedom initially available in $\phi^{aI}$. In this section we aim to deal with these remaining dofs. It is understood that the coset has the right structure to admit the tetrad as part of its subspace once the internal part of the homogenous coset has been projected out. This is done using a real unit internal vector field $n^{I}$, which is normalized according to $n^{I} n_{I}=1$:
\begin{equation}
\label{eq:tetrad-projection}
e^{a}(x)\;:=\phi^{a}{}_{I}(x)\,n^{I}(x)\,.
\end{equation}
Let us begin with the following ansatz:    
     \begin{equation}
  \phi^{aI}
  = \frac{1}{\ell}\,e^{a} n^{I}  +\tau^{aI},
  \label{eq:Phi-split}
\end{equation}
where the additional dof's are carried by the field $\tau^{aI}$. 
The ansatz for this construction is as follows.
We introduce mutually orthogonal projectors (recalling that $n^2=1$)
\begin{eqnarray}
 && P_{IJ}=n_{I}n_{J}, 
  \qquad 
  Q_{IJ}=\delta_{IJ}-n_{I}n_{J},
\end{eqnarray}
with \( P^{2}=P, Q^{2}=Q \) and \(PQ=0\). Then projecting both parallel and orthogonal to $n^{I}$ defines the orthogonal (coset) piece as
\begin{equation}
  \tau^{aI}\equiv Q^{IJ}\, \phi^{a}{}_{J},
  \qquad
  n_{I} \tau^{aI}=0 \label{eqconstt},
\end{equation}
with the number of independent components of the original field $\phi^{aI}$.
This construction is fully gauge covariant. 

With these preparations one can reformulate the Lagrangian introduced in the previous Section with the new coset ansatz in equation \eqref{eq:Phi-split}. The field strength components are then written as
\begin{eqnarray}
    \mathcal{F}^{ab}&=&R^{ab}-\frac{1}{\ell^2}e^a\wedge e^b+\tau^{a}{}_{I}\wedge \tau^{Ib} \label{eqn:tauexpans1}\\
\mathcal{F}^{IJ}&=&F^{IJ}+\frac{1}{\ell}(\tau^I{}_{a}\wedge e^a n^J-n^Ie_a\wedge \tau^{aJ})+\nonumber\\
    &&+\tau^{I}{}_a\wedge\tau^{aJ} \\
     \mathcal{F}^{aI} &=& \frac{1}{\ell}( T^{a} n^{I} +  e^{a}\wedge P^{I})+\nonumber\\
&&+d\tau^{aI}+\omega^a_{~c}\wedge\tau^{cI}+\tau^a_{~J}\wedge A^{JI}, \label{eqn:tauexpans3}
\end{eqnarray}
where we can see the additional $\tau^{aI}$--dependent terms compared to equations (\ref{eqfab}), (\ref{eqfij}) and (\ref{eqfai}).
It is then clear that this $\tau$ field will introduce indirect interaction terms between all the fundamental forces (bar gravity and weak forces as before). Nevertheless, one still recovers equation \eqref{fulldecomposition2} and thus the main theory remains the unchanged, aside from this interesting field ubiquitous in its interactions. We will explore its implications in future work.  

\section{Phenomenological Implications}
Having established the symmetry-breaking structure and field content of the model, we now turn to outline some elements of the broad observable consequences - a detailed exploration will be reserved for future work. The $\mathrm{SO}(1,9,\mathbb{C})$ framework generically predicts a variety of phenomenological effects, ranging from parity-sensitive gravitational couplings to possible imprints in the gauge and cosmological sectors. The cleanest of these may be the gravitational and gauge birefringence arising once the pseudoscalar $\phi$ acquires a vacuum expectation values.

At low energies the relevant parity-odd contribution to the effective action from above is:
\begin{equation}
S \supset \frac{\gamma}{32 \pi G} \int \phi\, R^-_{ab}\wedge R^{-ab},
\end{equation}
which, for $\phi=1+\delta\phi$, yields a Chern–Simons correction to Einstein–Cartan gravity. Following the methods outlines by Alexander and Yunes \cite{Alexander2006}, linearising around a torsionless FRW background gives modified wave equations for gravitational perturbations,
\begin{equation}
\Box~ h_{ij} \pm i\frac{\dot{\phi}}{M_{\rm Pl}}\,\epsilon_{ikl}\partial^k \dot{h}^l_j =0,
\end{equation}
with the sign distinguishing helicities. The resulting dispersion relation $
\omega_\pm^2=k^2 \pm k\,\frac{\dot{\phi}}{M_{\rm Pl}},
\label{eq:disp}
$
implies that left- and right-handed gravitational waves propagate with slightly different phase velocities. The accumulated phase shift over a propagation time $t$ is
\begin{equation}
\Delta\Phi = \int^t \!\frac{\dot{\phi}}{M_{\rm Pl}}\,{\rm d}t' ,
\end{equation}
producing a rotation of linear polarisation analogous to cosmic birefringence in axion electrodynamics~\cite{Carroll1990,Alexander2006}. Here, however, $\phi$ is intrinsic to the unified connection, controlling both parity breaking and the relative weighting of the chiral sectors.

Things proceed in exact analogy for the electromagnetic and Yang-Mills waves. After symmetry breaking, the parity-odd term in the unified action
\begin{equation}
    S \supset \frac{\gamma}{32 \pi G} \int \phi \mathcal{F}_{IJ} \wedge \mathcal{F}^{IJ},
\end{equation}
projects onto the U(1) component of SO(6) $\cong$ SU(4) as $\frac{\kappa_\gamma}{4}\int \phi F \wedge F$ where $F$ now refers to the EM field strength. Linearising Maxwell's field equations \cite{Alexander2006} then gives:
\begin{equation}
    A_\pm '' + (k^2 \pm k \kappa_\gamma \phi') A_\pm = 0,
\end{equation}
where $k$ is the wavenumber. Then the two circular polarisations can be seen to acquire opposite phase shifts and linearly polarised light experiences a small rotation precisely as in the gravitational case, and once again this is governed by the field $\phi$.

Beyond these birefringent and parity-odd effects, the unified curvature also encodes a hierarchy of vertex structures that couple gravity to the internal gauge fields. These can emerge by considering small perturbations of the metric against a flat Minkowski in the typical manner used to recover an Einstein-Maxwell vertex, but also through the expansion of the YM term as seen in Eqns (\ref{eqn:tauexpans1}) - (\ref{eqn:tauexpans3}) and the coupling with the Stelle-West compensator field fluctuations around the vev as seen in Eqns (\ref{interaction2}) - (\ref{endinteractions}). For example, if we once again consider the unified kinematic term $S \supset \frac{\alpha}{2}\int F_{AB} \wedge \star F^{AB}$ and insert
Eqns (\ref{eqn:tauexpans1}) - (\ref{eqn:tauexpans3}), retaining terms up to quadratic order in $\tau$, the relevant piece of the unified Lagrangian then reads:
\begin{align}
\mathcal{L}_\tau = \alpha &[ ~(D\tau)_{aI} \wedge * (D\tau)^{aI} + 2 \tau_{aI} \wedge \tau_{b}^{I} \wedge * R^{ab} \\ & + 2 (D\tau)_{aI} \wedge \phi^{aI} + 2 \tau_{aI} \wedge \tau^{a}_J \wedge * F^{IJ}~ ] \nonumber \\ &+ \frac{1}{2} m_\tau^2 \tau_{aI} \wedge *\tau^{aI}, \nonumber
\end{align}
where we have written $(D\tau)_{aI} = d\tau_{aI} + \omega^c_a \wedge \tau_{cI} + \tau_a^J \wedge A_{JI}$, $A_{JI}$ being the SO(6) gauge field. The mass term $m_\tau$ arises from internal symmetry breaking in the adjoint Higgs sector. Varying this with respect to $\tau_a^I$ yields the covariant field equation in a vacuum:
\begin{equation}
(D^2 + m_\tau^2) \tau_{aI} = -\alpha \left[ *R_{ab} e^b n_I + *F_{IJ} e_a n^J \right]. \label{taueom}
\end{equation}
As $m_\tau$ is set by the symmetry–breaking scale, these
fields are heavy and can be integrated out at tree level by solving
Eqn \eqref{taueom} perturbatively.
To leading order in $1/m_{\tau}^{2}$ one obtains the on–shell solution
\begin{equation}
\tau_{aI}\simeq
 -\frac{\alpha}{m_{\tau}^{2}}
   \Big(
     *R_{ab}\,e^{b}n_{I}
     + *F_{IJ}\,e_{a}n^{J}
   \Big),
\end{equation}
which, when substituted back into $\mathcal{L}_\tau$, gives us:
\begin{equation}
\mathcal{L}_\text{eff} \simeq \frac{c_1}{m_\tau^2} R_{\mu\nu\rho\sigma} F^{\mu\nu} F^{\rho\sigma} + \frac{c_2}{m_\tau^2} R_{\mu\nu} F^{\mu\alpha} F^{\nu}_{\alpha} + \frac{c_3}{m_\tau^2}RF^2,
\end{equation}
where the coefficients $c_i \sim \alpha^2$ encode geometric normalisations. The first term defines the graviton-gauge-gauge vertex, which contributes to the tree-level amplitude $h \to \gamma\gamma$ and analogous $h \to g g $, $h \to ZZ $  and $ h \to W^+ W^- $ processes. Of course, this term does not provide the only contribution to these vertices - this detailed calculation will be carried out in full detail in future work. This sector should be understood as an effective field theory description valid for energies $E \ll m_\tau$, at which the additional connection components become dynamical. Below this scale, $\tau$ can be integrated out, yielding the operators seen above. In this limit, the explicit mass term $\frac{1}{2}m_\tau^2$ simply parametrises the suppression by a factor of $\frac{1}{m_\tau^2}$ and does not need to be retained if $\tau$ is treated as auxiliary.

Note that all coefficients are fixed by the same geometric constants $(\alpha, \ell, m_{\tau})$ that determine $G$ and $\Lambda$. Thus, the SO(1,9) construction predicts a unified set of graviton-gauge vertices whose parity structure, coupling strengths, and suppression scales are not arbitrary but dictated by the symmetry-breaking dynamics of the graviweak sector \cite{AlexanderMarcianoSmolin2014, Nesti:0706.3307, AlexanderMarcianoTacchi2012}. Observable consequences may include chiral asymmetries in stochastic gravitation-wave spectra, polarisation rotation in electromagnetic and CMB signals, and suppressed but calculable curvature-driven corrections to gauge-boson scattering. These signatures offer falsifiable probes of the geometric unification mechanism proposed here.

\section{Conclusions and Outlook}
\label{conc}
We have presented a minimal graviGUT unification based on gauging $\mathrm{SO}(1,9,\mathbb{C})$ over four-dimensional spacetime, in which one of the Pati–Salam chiral $\mathrm{SU}(2)$ factors is geometrized as a chiral half of the Lorentz group rather than as an additional weak interaction. This identification exploits the decomposition $\mathfrak{so}(1,9)_{\mathbb{C}}\cong\mathfrak{so}(1,3)_{\mathbb{C}}\oplus\mathfrak{so}(6)_{\mathbb{C}}\oplus(\text{coset})$ and the bifundamental index structure of the coset, enabling a McDowell–Mansouri–type ansatz to solder the Lorentz and internal sectors through a single  field. In this setting a parity-symmetric chiral $\mathrm{SO}(1,9,\mathbb{C})$ action was constructed in which a pseudoscalar Higgs triggers dynamical parity breaking so that one chirality reduces to the self-dual gravitational sector and the opposite chirality becomes the Yang–Mills connection of the weak force. Matching to first-order Einstein–Cartan theory fixes the relative normalizations of the curvature terms and ties the internal parameters to observables, with the gravitational coupling and cosmological constant emerging from the relations among the $\mathrm{SO}(1,9,\mathbb{C})$ couplings.

We also analyzed the coset field $\phi^{a}{}_{I}$ and proposed the fate of its surplus degrees of freedom via the covariant split $\phi^{a}{}_{I}=\ell^{-1}e^{a}n_{I}+\tau^{a}{}_{I}$. This makes explicit how the tetrad arises while preserving full $\mathrm{SO}(6)$ covariance, and it isolates an orthogonal sector $\tau^{a}{}_{I}$. 
This field mediates a distinctive pattern of non-minimal interactions predicted by the unified curvature, offering novel phenomenological handles which will be revisited in future work. 

Furthermore, the unified origin of the kinetic terms implies specific, parity-sensitive non-minimal couplings among gravitons/weak bosons and the $\mathrm{SU}(3)\times\mathrm{U}(1)$ sector, together with the absence of a separate right-handed weak gauge boson—feature. Particularly promising are parity-sensitive gravitational-wave signatures and mixed curvature–gauge vertices whose strengths are fixed relative to $G$ and $\Lambda$ in our matching, providing falsifiable targets for upcoming probes.

Several directions follow naturally from this work. A good starting point is working out how to embed fermions and Yukawa structures in the $\mathrm{SO}(1,9,\mathbb{C})$ framework, check anomaly cancellation \cite{Witten1982,AlvarezGaumeWitten1984}, and study induced CP and flavor structures. Moreover, we can explore self-dual formulations, spin-foam/spinor variables \cite{FreidelKrasnov2008}, and topological terms compatible with the single-invariant origin of the low-energy theory. From a cosmological side, one can try to quantify parity-odd imprints in stochastic gravitational-wave backgrounds and study early-universe dynamics tied to the parity-breaking sector \cite{Contaldi:2008iw,Lue:1999rpc,Saito:0705.3701}. Finally, we can derive concrete amplitudes for mixed graviton–gauge processes and map them onto gravitational-wave constraints. Each item is algorithmic and testable within the present setup, reinforcing the central message: geometrizing one chiral $\mathrm{SU}(2)$ inside $\mathrm{SO}(1,9,\mathbb{C})$ yields a predictive and economical GraviGUT with crisp signatures that differentiate it from both traditional Pati–Salam and larger unified schemes.

It is perhaps telling that this line of reasoning—connecting Pati–Salam’s vision with the chiral geometric structures of quantum gravity—remained unexplored for decades (our paper could have been written in the early 1990s). Historical currents in the field often separated the grand-unification and quantum-gravity communities, the former favoring phenomenological pragmatism and the latter drifting toward ever more abstract formalisms. The present work reopens the dialogue between these seemingly discordant approaches, showing that a synthesis 
may in fact be the most natural route forward. In this sense, the recovery of a geometric Pati–Salam perspective within a chiral, self-dual gravitational framework is not merely technical: it is a small act of scientific reconciliation.

\acknowledgments
We thank Altay Etkin, Christopher Isham, Raymond Isichei, Loukas Gouskos, Kellogg Stelle, Subhash Wadwani and Daniel Waldram for helpful conversations. BA was supported by FCT Grant No. 2021.05694.BD and JM partly supported by STFC Consolidated Grant ST/T000791/1.


\twocolumngrid

\end{document}